\documentstyle[12pt,aasms4]{article}

\righthead{Galactic Gamma-Ray Background Radiation}
\lefthead{Berezhko \& V\"olk}
\newcommand{\gsim}{\,\raisebox{0.2em}{$>$}\!\!\!\!\!
\raisebox{-0.25em}{$\sim$}\,}
\newcommand{\lsim}{\,\raisebox{0.2em}{$<$}\!\!\!\!\!
\raisebox{-0.25em}{$\sim$}\,}
\newcommand{\gr}{$\gamma$-ray \,}
\newcommand{\grs}{$\gamma$-rays \,}
\begin{document}
\title{Galactic Gamma-Ray Background Radiation
          from Supernova Remnants}

\author{E.G. Berezhko}

\affil{Institute of Cosmophysical Research and Aeronomy, Lenin
Ave. 31, 677891 Yakutsk, Russia}
\authoremail{berezhko@sci.yakutia.ru}

\and

\author{H. J. V\"olk}

\affil{Max--Planck--Institut f\"ur Kernphysik, P.O. Box 103980,
D-69029 Heidelberg, Germany}
 \authoremail{Heinrich.Voelk@mpi-hd.mpg.de}


\begin{abstract} 
The contribution of the Source Cosmic Rays (SCRs), confined in Supernova
Remnants, to the diffuse high energy \gr emission above 1 GeV from the
Galactic disk is studied. \grs produced by the SCRs have a much harder
spectrum compared with those generated by the Galactic Cosmic Rays which
occupy a much larger residence volume uniformly. SCRs contribute less than
10\% at GeV energies and become dominant at \gr energies above $100$~GeV.
The contributions from $\pi^0$-decay and Inverse Compton \grs have
comparable magnitude and spectral shape, whereas the Bremsstrahlung
component is negligible. At TeV energies the contribution from SCRs
increases the expected diffuse \gr flux almost by an order of magnitude.
It is shown that for the inner Galaxy the discrepancy between the
observed
diffuse intensity and previous model predictions at energies above a few
GeV can be attributed to the SCR contribution.
\end{abstract}

\section{Introduction} 
Observations of the diffuse Galactic \gr emission give information about
the Galactic Cosmic Rays (GCRs), the interstellar gas and diffuse photon
fields, and about the interactions between them. The observational results
obtained with the Energetic Gamma Ray Experiment Telescope (EGRET) on the
Compton Gamma Ray Observatory can be described fairly well by a suitable
model for the diffuse interstellar gas, Cosmic Ray (CR) and photon
distributions (e.g. Hunter et al. 1997a; Hunter et al. 1997b).

However, above 1~GeV the observed average diffuse \gr intensity in the
inner Galaxy, $300^{\circ}<l<60^{\circ}$, $|b|\leq 10^{\circ}$, exceeds
the model prediction significantly. There are at least two possible
explanations for this discrepancy (e.g. Hunter et al. 1997b; Weekes et al.
1997). The high-energy $\gamma$-ray excess may indicate that the GCR
spectrum observed in the local neighborhood is not representative of the
diffuse CR population in the Galactic disk; a harder average diffuse
proton spectrum is required to explain the $\gamma$-ray excess if it is
due to $\pi^0$-decay. An unresolved distribution of CR sources is the
other possibility.

The physical picture which we consider in this paper corresponds to the
second possibility. The idea that CRs, after leaving their sources, could
in principle produce \grs in ambient dense clouds with a harder spectrum
than those produced by the average GCRs, was proposed by Aharonian and
Atoyan (1996).
Another proposed possibility, also invoking transport effects, is the
local hardening of the CR energy spectrum in the direction perpendicular to
the Galactic disk above strong CR sources, especially above the inner
region of the Galaxy, due to faster CR convection in a faster Galactic
Wind (V\"olk 1999).
%
%
This contributes a principally observable harder than average \gr
component in such regions. In contrast, we consider here the accelerating
particles inside their sources, where they are much more strongly
scattered than in the ISM, neglecting the contributions invoked by
Aharonian and Atoyan, and by V\"olk (see also below).

We assume that SNRs are the dominant sources of the Galactic CRs. On this
premise we find that CRs, accelerated and confined in SNRs, give an
important contribution to the high-energy $\gamma$-ray emission from the
Galactic disk. Since the CR energy spectrum inside SNRs is much harder
than on average in the Galaxy --- the average spectrum being softened by
rigidity-dependent escape from the Galaxy in the diffusion region above
the disk --- the relative SNR contribution increases with energy and becomes
in fact dominant at \gr energies $\epsilon_{\gamma} \gsim 100$~GeV. It may
substantially increase the diffuse TeV $\gamma$-ray emission from the
Galactic disk so as to constitute a significant and spatially variable
observational background which must also be taken into account in the
search for spatially extended Galactic CR sources in this energy region.
A physically analogous problem is the contribution of CR electrons in SNRs
to the {\it radio synchrotron spectrum} of normal galaxies, without an
Active Galactic Nucleus. It has been recently discussed by Lisenfeld \&
V\"olk (1999).

In this paper we shall investigate pion-decay $\gamma$-ray emission from
CR nuclei as well as Inverse Compton (IC) radiation and Bremsstrahlung due
to CR electrons. It will be shown that the average IC $\gamma$-ray
background from SNRs is comparable in magnitude and spectral form to the
pion-decay background at high energies, whereas the corresponding
Bremsstrahlung component is negligible.

\section{Gamma-ray luminosity of old SNRs}
The majority of the GCRs, at least up to kinetic energies $\epsilon \sim
10^{14}$~eV, is presumably accelerated in SNRs. According to modern theory
a significant part of the hydrodynamic Supernova (SN) explosion energy
$E_{SN}\sim 10^{51}$~erg is converted into CRs already in the early Sedov
phase of the evolution, due to diffusive shock acceleration (e.g. Berezhko
et al. 1996; Berezhko \& V\"olk 1997). Later on, the CR energy content and
the high-energy $\gamma$-ray production slowly decrease with time. This is
at least true as long as the progenitor star is not so massive as to have
a strong wind which significantly modifies the circumstellar medium
(Berezhko \& V\"olk, 2000). 
The total number of SNRs
$N_{SN}=\nu_{SN}T_{SN}$ is an increasing function of their assumed life
time $T_{SN}$, i.e. the time until which they can confine the accelerated
particles; here $\nu_{SN}$ is the Galactic SN rate. Therefore we conclude
that the population of the oldest SNRs dominates the total $\gamma$-ray
luminosity of the ensemble of Galactic SNRs. Thus we consider only old
SNRs which nevertheless are still strong enough to confine most of the CRs
produced during the prior evolutionary stages.

We then have the situation that the CRs in the Galaxy are represented by
two basically different populations. The first one consists of the
ordinary GCRs and presumably occupies a large Galactic residence volume
quasi-uniformly. This residence volume exceeds that occupied by the CR
sources by far (e.g. Ptuskin et al. 1997; for an earlier review, see
Berezinsky et al. 1990). The second CR population, which we call Source
Cosmic Rays (SCRs), is represented by shock accelerated CRs that are still
confined in the localized SNRs. During the initial, active period of SNR
evolution of about $t\lsim10^5$~yr when the SN shock is relatively strong,
the volume occupied by the accelerated CRs practically coincides with the
shock volume. In later stages the shock becomes weak and CRs begin to
leave the SNR acceleration region. After some period of time $T_{SN}$ the
escaping SCRs become very well mixed with the ``sea'' of GCRs. We shall
assume that the transitory period during which SCRs are transformed into
GCRs is much shorter than the preceding confinement time $T_{SN}$. The
most important factor for CR confinement is the shock strength. Even in
the phase where radiative cooling would formally become important, this
remains true since CRs prevent cooling compression due to their pressure.
Thus the assumed mean life time $T_{SN}\lsim10^5$~yr is determined by the shock
dynamics more than by anything else.

Since the $\gamma$-ray production due to GCRs is quite well studied (e.g.
Hunter et al. 1997b; Mori 1997), it is primarily important to find the
relative contribution of the SCR population.

\subsection{Gamma rays from $\pi^0$-decay}
The production rate of $\pi^0$-decay \grs from inelastic CR - gas
collisions, primarily p-p collisions, may be written in the form (Drury et
al. 1994)
\begin{equation}
Q_{\gamma}(\epsilon)=Z_{\gamma}\sigma_{pp} c N_g n(\epsilon),
\end{equation} 
where $N_g$ is the local gas number density, $\sigma_{pp}$
is the inelastic p-p cross-section, $Z_{\gamma}$ is the so-called
spectrum-weighted moment of the inelastic cross-section,
$n(\epsilon)d\epsilon $ is the CR spatial number density of CRs in the
kinetic energy interval $d\epsilon$, and $c$ is the speed of light. Thus
we have to primarily calculate $n(\epsilon)$ for the two CR populations.

The quasi-uniform GCR population in the gas disk is
assumed to have roughly a power law spectrum in the relativistic range
\begin{equation} n_{GCR}(\epsilon)=\frac{n_0^{GCR}(\gamma_{GCR}-1)}{mc^2}
 \left( \frac{\epsilon}{mc^2}\right)^{-\gamma_{GCR}}.
\end{equation}
The total number $n_0^{GCR}$ of relativistic GCRs with $\epsilon>mc^2$,
per unit volume, can be expressed in terms of the CR energy density
$e_{GCR}$:
\begin{equation}
n_0^{GCR}=\frac{(\gamma_{GCR} -2)e_{GCR}}{(\gamma_{GCR}-1)mc^2}, 
\end{equation} 
where $m$ is the proton mass. For simplicity we restrict our consideration
here to the proton component which is energetically dominant in both the
GCR and the SCR populations.

In contrast to the GCR population, the SCRs are confined inside a discrete
number $N_{SN}$ of SNRs. These are assumed to be predominantly located in
the Galactic gas disk, of volume $V_g$. Spatially averaged over the
Galactic disk volume their $\gamma$- ray production rate is determined by
an expression analogous to eq. (1), where instead of $n(\epsilon)$
one should substitute the SCR distribution
\begin{equation}
n_{SCR}(\epsilon)=N_{SCR}(\epsilon)N_{SN}/V_g,
\end{equation}
with $N_{SCR}(\epsilon)d\epsilon$ being the overall (i.e. integrated over
the SNR volume) SCR number in the energy interval $d\epsilon$, and should
use an appropriate mean local interstellar medium (ISM) number density
$N_g^{SCR}$ into which the SNe explode.

Since the CRs produced inside SNRs have also a power-law spectrum
$N_{SCR}\propto \epsilon^{-\gamma_{SCR}}$ in the relativistic range,
$n_{SCR}(\epsilon)$ can be expressed in the same forms
 (2) and(3), with
\begin{equation}
e_{SCR}=N_{SN}\delta E_{SN}/V_g, 
\end{equation} 
and putting $\gamma_{SCR}>2$. Here $\delta$ is the fraction of the SN
explosion energy $E_{SN}$ converted into SCRs.

However, according to the prediction from nonlinear kinetic theory
(Berezhko et al. 1996), diffusive shock acceleration produces an extremely
hard spectrum of SCRs at the early Sedov phase which is characterized by a
power law index $\gamma_{SCR}=2$. In this case we have
\begin{equation} n_0^{SCR}=\frac{e_{SCR}}{mc^2\ln(\epsilon_{max}/mc^2)}, 
\end{equation} 
where $\epsilon_{max}$ is the maximum SCR energy.

For the ratio $R=Q_{\gamma}^{SCR}/Q_{\gamma}^{GCR}$ of the $\gamma$-ray
production rates due to SCRs and GCRs, we have
\begin{eqnarray} R(\epsilon_{\gamma})&=&
\frac{Z_{\gamma}^{SCR} N_{SN}\delta E_{SN}}
{Z_{\gamma}^{GCR}(\gamma_{GCR}-2) \ln
(\epsilon_{max}/mc^2) V_g e_{GCR}}\nonumber \\
             & \times &
\zeta \left(\frac{\epsilon_{\gamma}}{mc^2}\right)^{\gamma_{GCR}-2},
\end{eqnarray}
where $\zeta$ is the ratio $N_g^{SCR}/N_g^{GCR}$, and $N_g^{GCR}$ denotes
the average gas density in the disk. In fact we assume that the gas and
the CRs are distributed uniformly inside each SNR which is approximately
true for the old SNRs which we consider here. The parameter $\zeta$
describes a possible spatial correlation between SN occurrence and local
ISM density. If on average Supernovae explode in a denser than average
medium in the Galactic disk, then $\zeta >1$, whereas $\zeta<1$ in the
opposite case.

The main mass of the ISM $M_g=4\times 10^9M_{\odot}$ is contained in a
Galactic disk region of a thickness of about 240~pc, corresponding to the
thickness of the HI gas (Dickey and Lockman 1990) which has the volume
$V_g=2.5\times 10^{66}$~cm$^3$ referred to before. Here we take a disk
radius of about $10$ kpc which implies an average gas density
$N_g^{GCR}=2~
{\rm cm}^{-3}$. Taking the relativistic part of the GCR spectrum to be
characterized by $e_{GCR} \simeq 10^{-12}$~erg/cm$^3$, and
$\gamma_{GCR}=2.75$ which results in
$Z_{\gamma}^{SCR}/Z_{\gamma}^{GCR}=10$ (Drury et al. 1994), we obtain for
the standard set of SN parameters $E_{SN}=10^{51}$~erg,
$\nu_{SN}=1/30$~yr$^{-1}$:
\begin{equation} R(\epsilon_{\gamma})=0.16
\zeta \left(\frac{T_{SN}}{10^5~\mbox{yr}}\right)
\left(\frac{\epsilon_{\gamma}}{1~\mbox{GeV}}\right)^{0.75}, 
\end{equation}
in addition using the rather moderate parameter values $\delta =0.1$ and
$\epsilon_{max}=10^5 \, mc^2$ to characterize CR acceleration inside SNRs
(e.g. Berezhko et al. 1996). One can see from this expression that for
$T_{SN}\sim 10^5$~yr the $\gamma$ ray production due to SCRs becomes
dominant
already at energies $\epsilon_{\gamma} \gsim 10$~GeV.

Note that the quantity $\delta E_{SN}N_{SN}/(V_ge_{GCR})$ represents the ratio
of currently existing total SCR energy and GCR energy inside $V_g$. For
the above set of parameters it is about 0.1. Despite the fact that the
SCRs represent only a relatively small fraction of the total CR energy
content even in the disk, they may dominate the \gr production at
sufficiently high energies due to their much harder spectrum.

It is clear that the quantity $R(\epsilon_{\gamma})$ determines the
average ratio \\
$(dN_{\gamma}^{SCR}/d\epsilon_{\gamma})/(dN_{\gamma}^{GCR}/d\epsilon_{\gamma})$
of \gr spectra produced in any region of the disk, by SCRs and GCRs,
respectively.  Therefore the total \gr spectrum measured from an arbitrary
Galactic disk volume is expected to be 
\begin{equation} {dN_{\gamma}
\over d\epsilon_{\gamma}}= {dN_{\gamma}^{GCR}\over d\epsilon_{\gamma}}
[1.4+R(\epsilon_{\gamma})], 
\end{equation} 
where the additional factor
0.4 is introduced to approximately take into account the contribution of
GCR electron component to the diffuse \gr emission at GeV energies (e.g.
Hunter et al. 1997b). In Fig.1 we present the expected differential flux
of \grs from the inner Galaxy, calculated for $\zeta=1$ and
$T_{SN}=10^5$~yr, with the spectrum $dN_{\gamma}^{GCR}/d\epsilon_{\gamma}$
taken from the paper by Hunter et al. (1997b), and extended into the
region $\epsilon_{\gamma}> 30$~GeV according to the law
$\epsilon_{\gamma}^{-2.75}$. One can see that after inclusion of the \grs
produced by SCRs, the calculated flux even exceeds the EGRET flux for
$\epsilon_{\gamma}\gsim 20$~GeV. This suggests that expression (7)
overestimates the \gr production by SCRs.

It is possible that the overall source spectral index $\gamma_{SCR}$ is
somewhat larger than 2 due to very late accumulation of only low-energy
particles. It can also not be excluded that the confinement time $T_{SN}$
of the SCRs depends on energy. Due to their high mobility, the highest
energy particles may leave the vicinity of parent SNR earlier and also
more rapidly. This process of SCR escape into the ISM starts for the most
energetic particles already at the early Sedov phase of SNR evolution
(e.g. Berezhko et al. 1996; Berezhko \& V\"olk 1997). Therefore at time
$T_{SN}$, when the main part of SCRs are released from the SNR, their
spectrum may be somewhat steeper than a $\gamma_{SCR}=2$ spectrum.


Due to the importance of the problem we derive the relation between
$n_{SCR}$ and $n_{GCR}$ in a different form. It leads to the same results
if SNRs are the GCR source. We start from the usual leaky box balance
equation
\begin{equation} \frac{n_{GCR}(\epsilon)}{\tau_c}=
\frac{N_{SCR}}{V_c}\nu_{SN}, 
\end{equation} 
where $V_c(\epsilon)$ is the energy-dependent residence volume occupied by
GCRs that reach the gas disk during their mean residence time $\tau_c$
in $V_c$ . In the case of an
extended Galactic halo due to a Galactic wind driven by the GCRs
themselves, and
for energies much larger than a few GeV, $V_c$ can be much
greater than $V_g$. In fact  $V_c(\epsilon)\propto \epsilon^{0.55}$ in such a
selfconsistent halo model (Ptuskin at al. 1997). 

Note that the leaky box model deals with a CR distribution
$n_{GCR}(\epsilon)$ averaged over the residence volume $V_c$. Therefore it
can only be applied to the GCRs which can be assumed to be almost
uniformly
distributed in the residence volume. It is not valid for the SCRs, because
their behavior is determined not only by large-scale transport but also by
other physical factors which, for example, lead to their acceleration.
Technically the volume $V_{SCR}$, occupied by the SCRs, should be
excluded from the residence volume $V_c$ and the SCRs appear in the
balance equation (10) for the GCRs only in the form of a source term
$N_{SCR}\nu_{SN}/V_c$ as a CR population released from the source region
into the ISM after some unspecified evolutionary period $T_{SN}$.
Therefore eq.(10) does not depend upon $T_{SN}$. However,
the effects produced by the SCRs confined inside the ensemble of
simultaneously existing SNRs, for example the additional \gr production,
essentially depends on $T_{SN}$, since it directly determines the total
number $N_{SN}$ of simultaneously existing SNRs.

Using eq. (4) we can write 
\begin{equation}
\frac{n_{SCR}}{n_{GCR}}=\frac{V_cT_{SN}}{V_g\tau_c} =
\frac{T_{SN}}{\tau_g}. 
\end{equation} 
The GCR residence time in the disk
volume, $\tau_g=\tau_c V_g/V_c$, can be derived from the measured grammage
$x$, which is the mean mass of Interstellar matter traversed by GCRs of
speed $v$ in the course of their random walk in the Galaxy:
\begin{equation}
\tau_g=\frac{x V_g}{v M_g}.
\end{equation}
The measured grammage at high energies $\epsilon\geq \epsilon_0=4.4$~GeV
is \\
$x=14\, (v/c)\, (\epsilon/4.4~\mbox{GeV})^{-0.60}$~g/cm$^2$; for $\epsilon<
\epsilon_0$, $x=14 v/c$~g/cm$^2$ (Engelman et al. 1990). Therefore
the
residence time {\it in the gas disk} can be written in the form
\begin{equation}
\tau_g = \tau_0 (\epsilon /\epsilon_0)^{-\mu},
\end{equation}
where $\tau_0 =4.6\times 10^6$~yr, $\epsilon_0=4.4$~GeV, $\mu=0.6$, and
$\epsilon\geq \epsilon_0$. This experimentally inferred value for $\mu$
closely agrees with the theoretical result of Ptuskin et al. (1997).

We note that at relativistic energies $\epsilon > mc^2$, according to the
initial balance eq.(10), the GCR spectrum $n_{GCR}\propto
\epsilon^{-\gamma_{GCR}}$ and the overall SCR spectrum $N_{SCR}\propto
\epsilon^{-\gamma_{SCR}}$ should be connected by the relation
\begin{equation} \gamma_{SCR}=\gamma_{GCR}-\mu.
\end{equation} 
For $\mu =0.60$
the source should produce a SCR spectrum with $\gamma_{SCR}=2.15$, while
for the case $\gamma_{SCR}=2$ one would need $\mu=0.75$.

At the same time the SCR distribution \\ $n_{SCR}(\epsilon)
\propto\epsilon^{-\gamma_{SCR}'}$, averaged over the gas disk, can have a
different shape compared to
$N_{SCR}(\epsilon)$ in the case of an energy dependent SNR confinement
time
$T_{SN}(\epsilon)$, according to eq. (11). It can be steeper,
$\gamma_{SCR}'=\gamma_{GCR}-\mu+\beta$, if $T_{SN}\propto
\epsilon^{-\beta}$
is a decreasing function of energy. It would mean that the highest energy
SCRs
 leave the parent SNR faster than the lower energy SCRs.

Eq.(11) leads to a simple expression for the \gr production ratio
\begin{equation}
R(\epsilon_{\gamma})=\zeta
\frac{Z_{\gamma}^{SCR} T_{SN}}{Z_{\gamma}^{GCR}\tau_g},
\end{equation}
independently of $\nu_{SN}$. Substituting the residence time in the form
(13), and taking $\gamma_{SCR}=2.15$, which leads to
$Z_{\gamma}^{SCR}/Z_{\gamma}^{GCR}=7.5$ (Drury et al. 1994) in eq.(15), we
obtain for $\epsilon_{\gamma}\geq 4.4$~GeV:
\begin{equation}
R(\epsilon_{\gamma})=0.07 \zeta
\left(\frac{T_{SN}}{10^5~\mbox{yr}}\right)
\left(\frac{\epsilon_{\gamma}}{1~\mbox{GeV}}\right)^{0.6}.
\end{equation}

We shall only consider \gr energies $\epsilon_{\gamma}\geq 4.4$~GeV. One can see
here again that the SCR contribution is determined by the value of the
confinement time inside SNRs, $T_{SN}$. Unfortunately there is no detailed
description of when and how CRs, accelerated in SNR, are released into the
ISM. Nevertheless one can give some constraints  on this process.

According to the standard theory, the expanding SNR shock produces a power
law CR spectrum up to the maximum energy (Berezhko 1996; Berezhko et al.
1996; Berezhko \& V\"olk 1997)
\begin{equation}
\epsilon_m\propto R_sV_s,
\end{equation}
which is determined by the radius $R_s$ and speed $V_s$ of the shock. The
CRs with the highest energy $\epsilon_{max}$ are produced at the very
beginning of the Sedov phase $t\sim t_0$ when the product $R_sV_s$ has its
maximum $R_0V_0$, where
\begin{equation}
t_0=\frac{R_0}{V_0},~ R_0=\left( \frac{3M_{ej}}{4\pi
\rho_g}\right)^{\frac{1}{3}},~ V_0=\sqrt{\frac{2E_{sn}}{M_{ej}}}
\end{equation} 
are the sweep-up time, sweep-up radius and initial mean ejecta speed
respectively; $M_{ej}$ denotes the ejecta mass, and $\rho_g=N_gm$ the ISM
density. Subsequently, the product $R_sV_s$ decreases with time as
$t^{-1/5}$ and the SNR shock produces CRs with progressively lower cutoff
energy $\epsilon_m(t)<\epsilon_{max}=\epsilon_m(t_0)$. During that phase those 
CRs that were previously produced with energies
$\epsilon_m<\epsilon<\epsilon_{max}$ now propagate outward diffusively
without significant influence of the SNR shock. If their expansion is
still governed by the Bohm diffusion coefficient as during their
acceleration, the expansion rate is only slightly higher than the SNR
expansion rate and these particles should be considered as confined inside
the source (i.e. the SNR). 

The opposite extreme case corresponds to the assumption that particles
with energies $\epsilon >\epsilon_m$ do no more produce a high level of
turbulence. Let us consider this pessimistic scenario in terms of
confinement here. In this case the propagation of these very high energy
particles is governed by the mean Galactic diffusion coefficient which is
very much larger than the Bohm diffusion coefficient. In this situation
particles with energy $\epsilon$ should be considered as released from the
source at the moment $t\geq t_0$ when $\epsilon_m(t)$ drops below
$\epsilon$. Since the particles with maximum energy are produced at
$t\approx t_0$, one can write
\begin{equation}
T_{SN}(\epsilon)=t_0\left( \frac{\epsilon}{\epsilon_{max}}\right)^{-5}.
\end{equation} 
Due to this strong dependence it is clear that $T_{SN}(\epsilon)$ will
still deviate from the overall gas dynamic life time $T_{SN}^{tot}$ only
for large energies $\epsilon$ near $\epsilon_{max}$. All particles with
%
$$\epsilon/\epsilon_{max} < (T_{SN}^{tot}/t_0)^{-1/5}$$
%
will remain confined until $T_{SN}^{tot}$. 

Since the majority of Galactic SNe are core collapse SNe from stars with
masses exceeding about $8 M_{\odot}$, we shall use $M_{ej}=10\, M_{\odot}$
for purposes of estimate. Except for progenitor masses exceeding $15\,
M_{\odot}$, the progenitors have only a weak stellar wind. Therefore the
assumption of a uniform circumstellar medium remains reasonable for the
average properties of the SNR population.

For the main fraction of CRs the confinement in SNRs terminates when the
SNR shock becomes weak and produces CRs with a very steep spectrum that
cannot anymore excite a high level of turbulence near the shock front. If
we take $M=4$ as a critical Mach number, the corresponding SNR age will be
$t=(M_0/M)^{5/3}=1.5\times10^3\,t_0$, where $M_0=V_0/c_{S0}$ is the
initial shock Mach number and $c_{S0}$ is the ISM sound speed. For an ISM
with number density $N_g=1$~cm$^{-3}$, $c_{S0}\simeq 4~$km/sec, and then
$t_0\simeq1.5\times 10^3$~yr which gives $T_{SN}\simeq 2\times 10^6$~yr
for $M_{ej}=10~M_{\odot}$ and $E_{SN}=10^{51}$ erg. This estimate shows
that the SNR shock remains rather strong during a very long period of
time.

Another physical factor which can restrict the SCR confinement is the
radiative cooling of the postshock gas. Approximately it becomes important
when the postshock temperature drops below $\sim 10^6$~K, or when the
postshock sound speed drops below $c_{S2}\sim 100$~km/s. In the case of a
strong shock, with Mach number $M\gg 1$, the postshock sound speed is
determined by the shock speed $c_{S2}\sim \sqrt{5}V_s/3$. During the Sedov
phase the shock speed decreases with time according to the law $V_s=0.4
\times V_0~(t/t_0)^{-3/5}$. Therefore the shock speed drops to the value
$V_c=100$~km/s at the age $t_c=t_0\,(0.4\,V_0/V_c)^{5/3}$. The above set
of SN and ISM parameters gives $t_c\sim 6\times 10^4$~yr. Since gas
clumping as a result of cooling may lead to effective SCR leakage from the
SNR, a value of the confinement time $T_{SN}=10^5$~yr is reasonable for
$N_g=1~ {\rm cm}^{-3}$. (Fig. 1).

\begin{figure}
\plotone{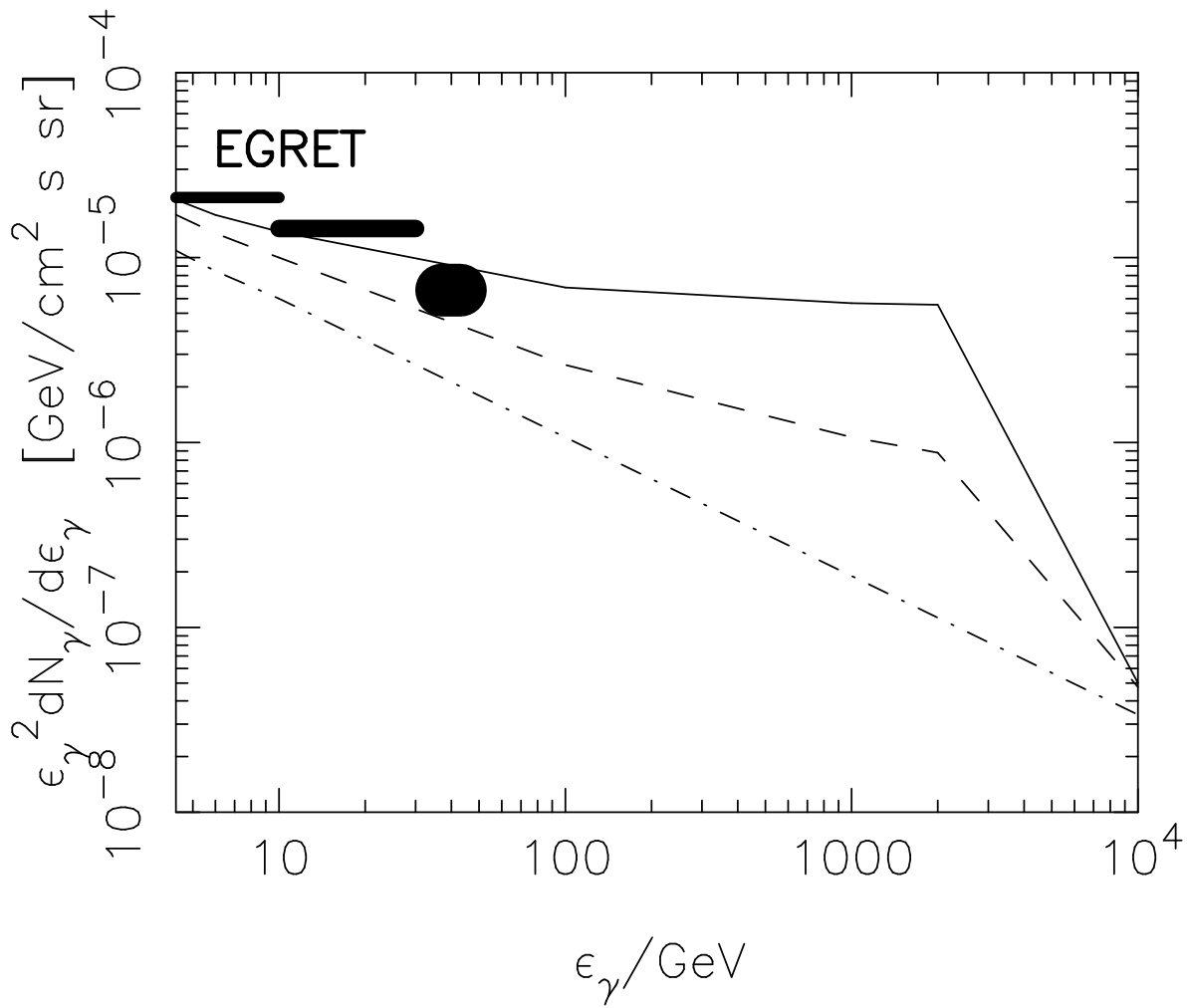}
\caption{
The average diffuse \gr spectrum of inner Galaxy
($300^{\circ}<l<60^{\circ}$, $|b|\leq10^{\circ}$). The full (dashed) line represents the
calculation with SCR spectral index $\gamma_{SCR}=2$
($\gamma_{SCR}=2.15$), and the
dash-dotted line corresponds to the $\pi^0$ decay \gr spectrum produced by
GCRs (Hunter et al. 1997b). EGRET data are also taken from the review 
paper by Hunter et al. (1997b).
\label{fig1} }
\end{figure}

\placefigure{fig1}

In Fig.1 we present a calculated \gr spectrum based on the above
expression (16) with 
\begin{equation}
T_{SN}=min\{10^5,10^3(\epsilon/\epsilon_{max})^{-5}\}~\mbox{yr},
\end{equation} 
$\epsilon_{max}=10^5$~GeV, and $\epsilon_{\gamma}=0.1\epsilon$,
which is roughly valid for the hadronic considered \gr
production process considered. 
At GeV energies in this case SCRs contribute
about 10\% of the total \gr flux. Due to their hard spectrum this
contribution progressively increases with energy and becomes dominant at
$\epsilon_{\gamma}\gsim 100$~GeV. It increases the expected TeV \gr flux
by about a factor of ten. Note that the actual SCR contribution from the
inner part of the Galaxy is somewhat higher than the above estimate due to
the larger SNR concentration in this region.

As one can see from Fig. 1, the SCR contribution in the case
$\gamma_{SCR}=2$ , using eqs. (8) and (9), is for all energies larger than
that for $\gamma_{SCR}=2.15$, using eq. (16). This difference is related
to the different SCR acceleration efficiencies. In the first case it is
characterized by the parameter $\delta =0.1$ which,  according to eq. (5),
directly determines the \gr production rate. In the second case the SCR
acceleration efficiency $\delta=[\int_{mc^2}^{\epsilon_{max}} \epsilon
N_{SCR}(\epsilon) \, d\epsilon]/E_{SN}$
is not contained in the final expression
(15), but one can derive it easily from the balance equation (10) which
gives: 
\begin{equation}
\delta \nu_{SN} E_{SN} =\int_{mc^2}^{\epsilon_{max}}
\frac{n_{GCR}V_g}{\tau_g}\epsilon d\epsilon.
\end{equation} 
To determine
$\delta$ we use the (demodulated) GCR distribution
\begin{eqnarray} n_{GCR}(\epsilon)&=& 8.1\times 10^{-10}\nonumber  \\
              & \times &
\left(\frac{\epsilon}{1~\rm{GeV}}+\frac{mc^2}{1~\rm{GeV}} 
\right)^{-2.75}~{\rm cm}^{-3}~{\rm (GeV)}^{-1}.
\end{eqnarray}
(Ryan et al. 1972; Perko 1987).

Substituting the values of $\tau_g(\epsilon)$, $V_g$, $E_{SN}$, and
$\nu_{SN}$, we obtain $\delta \simeq 0.05$ for the relativistic part of
the GCR spectrum. The required acceleration efficiency is two times lower
compared with the case $\gamma_{SCR}=2$ due to the essentially steeper
spectrum. Note, that in both cases the SCR spectra with $\gamma_{SCR}=2$,
$\delta=0.1$ and $\gamma_{SCR}=2.15$, $\delta=0.05$ contain about the same
number of relativistic CRs. In the first case the required GCR residence
time is $\tau_g\propto \epsilon^{-0.75}$, whereas the second case with
$\tau_g\propto \epsilon^{-0.6}$ is close to the experiment
(Engelman et al. 1990). Therefore we believe that the dashed line in Fig.1
represents the most reliable estimate for the expected diffuse
$\pi^0$-decay \gr emission, especially at high energies
$\epsilon_{\gamma}\gsim 100$~GeV.

Note also that the acceleration efficiency required by eq. (21),
using an assumed Galactic SN rate $\nu_{SN}=1/30$~yr$^{-1}$ and a mean SN
explosion energy $E_{SN}~=~10^{51}$~erg, is considerably lower than that
predicted by shock acceleration theory, which gives $\delta = 0.2\div 0.5$
(Berezhko et al. 1996). Yet, in contrast to the acceleration models which
determine $\delta$ from the injection rate and the nonlinear acceleration
theory selfconsistently, eq. (21) determines only the product $\delta
E_{SN} \nu_{SN}$ from observed quantities. The observationally inferred SN
explosion energies $E_{SN}$ can be at least by a factor 2 smaller than
$10^{51}$ erg, and from comparisons with galaxies similar to our own
$\nu_{SN}$ could vary between $1/30\div 1/100$~yr$^{-1}$. Therefore the
empirical value of $\delta$ from eq. (21) can vary between 0.05 and 1/3.
Nevertheless, the theoretically determined efficiencies appear
systematically too high.

As a possible solution for this discrepancy one might assume that, just
before being released, the SCRs lose an important part of their energy by
adiabatic expansion so that the SCRs' energy content inside a SNR is
higher than the energy contained in the released spectrum
$N_{SCR}(\epsilon)$. However, there is little dynamical basis for such an
assumption. Much more likely is that the very efficient CR acceleration
inside SNRs predicted by the nonlinear kinetic theory, assuming spherical
symmetry, takes place in reality only at some fraction of the SN shock
surface, because suprathermal positive ion injection into the acceleration
process on the highly oblique part of the shock can be significantly
suppressed (Bennet \& Ellison 1995; Malkov \& V\"olk 1995). In this case
the acceleration efficiency, calculated for a spherical SNR shock, should
be reduced by a factor of a few.

The actual SNR distribution can in fact be rather nonuniform within the
disk volume, contrary to what we have assumed implicitly up to now. In
this case the estimated value of $R(\epsilon_{\gamma})$, which describes
the relative SCR contribution to the diffuse \gr emission, should be
corrected by a factor $N_{SN}^a/N_{SN}$, where $N_{SN}^a$ is the
expected number of SNRs in the observed region and $N_{SN}$ represents
this number in the case of uniformly distributed SNRs. It is clear that
the expected value of $R(\epsilon_{\gamma})$ is almost independent of
the actual SNR distribution if the observed region is an essential part of 
the whole disk volume $V_g$.

\subsection{Inverse Compton and Bremsstrahlung gamma-rays from SCR
electrons}
Electrons, once being injected into the diffusive shock acceleration
process, will be as efficiently accelerated in SNRs as are the
protons. Even though there exist theoretical concepts (e.g. Levinson 1994;
Galeev et al. 1995; McClements et al. 1997; Bykov \& Uvarov 1999), 
electron injection is not
completely understood. However, there is no doubt that electrons undergo
continuous acceleration during SNR evolution. The spectral shape of
accelerated electrons $N_{SCR}^e(\epsilon)$ inside SNRs deduced from
radio-observations on average agrees with what is expected from shock
acceleration.
Since relativistic electrons with energies $\epsilon >1$~GeV are
dynamically indistinguishable from protons, their source spectrum
$N_{SCR}^e(\epsilon)=K_{ep}N_{SCR}(\epsilon)$ can differ from that of the
protons $N_{SCR}(\epsilon)$ only by some energy independent factor
$K_{ep}$ that is determined by the injection process. High energy
electrons produce \gr emission due to IC scattering, especially on the
Cosmic Microwave Background (CMB) and by Bremsstrahlung on the
interstellar gas. We shall first consider the IC contribution here.

\subsubsection{Inverse Compton contribution} 
In an approximate form, valid if the generating electron energy
distribution is smoothly varying, like in the case of a power law
considered here, the IC \gr emissivity
$Q_{\gamma}^{IC}(\epsilon_{\gamma})$ can be written as (e.g. Longair
1981, Berezinsky et al. 1990)
\begin{equation} 
Q_{\gamma}^{IC}(\epsilon_{\gamma})= \sigma_T c N_{ph} n^e_{SCR}(\epsilon_e)
\frac{d\epsilon_e}{d\epsilon_{\gamma}}, 
\end{equation}
where
\begin{equation}
\epsilon_e=m_e
c^2\sqrt{3\epsilon_{\gamma}/(4\epsilon_{ph})}
\end{equation}
is the energy of electrons
which produce an IC photon with mean energy $\epsilon_{\gamma}$,
$\sigma_T=6.65\times10^{-25}~{\rm cm}^2$ denotes the Thomson cross section,
$\epsilon_{ph}=6.7\times 10^{-4}$~eV and $N_{ph}=400$~cm$^{-3}$ are the
mean energy and number density of the CMB photons, respectively. Finally
$n^e_{SCR}=N^e_{SCR}N^e_{SN}/V_g$ denotes the
average spatial electron SCR number density.

Thus the ratio of the IC to the $\pi^0$-decay \gr production rate reads as
\begin{equation} 
\frac{Q_{\gamma}^{IC}}{Q_{\gamma}^{SCR}} = 1028 K_{ep}
\left(\frac{1 ~{\rm cm}^{-3}}{N_g^{SCR}}\right)
\left(\frac{ \epsilon_{\gamma}} {1~\mbox{TeV}} \right)^{1/2} 
\frac{N_{SN}^e}{N_{SN}} , 
\end{equation}
where we have used $\gamma_{SCR}=2$ and $\sigma_{pp}=4\times
10^{-26}$~cm$^{-2}$; $N_{SN}^e$ denotes the number of SNRs which
contribute to the IC emission at energy $\epsilon_{\gamma}$ from CR
electron sources.

From this expression it appears as if the IC \gr contribution would
be dominant at TeV-energies if $K_{ep}$ is as large as $10^{-2}$ and
if the mean gas number density inside SNRs is about $N_g^{SCR}=1$~cm$^{-3}$.
However, the radiative cooling time $\tau_e(\epsilon_e)$ of electrons,
which produce \grs with energy $\epsilon_{\gamma}=
(\epsilon_e/17.1~\mbox{TeV})^2$
~TeV, 
\begin{equation}
\tau_e= 7.3 \times 10^3 \left(\frac{10 ~\mu \mbox{G}}{B}\right)^{2}
\left(\frac{1~{\rm TeV}}{\epsilon_{\gamma}}\right)^{1/2}~\mbox{yr},
\end{equation}
reaches the above assumed overall SNR confinement time $T_{SN}=10^5$~yr
for \gr energies $\epsilon_{\gamma}<\epsilon_{\gamma}^{*} = 5.4 (B/10~ \mu
{\rm G})^{-4}$~GeV. Here $B$ is the magnetic field strength inside the
source, whose
typical value inside SNRs is about 10~$\mu$G. Therefore, taking into
account the obvious relation $N_{SN}^e/N_{SN}=\tau_e/T_{SN}$, on average
the relative IC contribution of the electron component of SCRs in TeV \gr
can be written as
\begin{equation}
\frac{Q_{\gamma}^{IC}}{Q_{\gamma}^{SCR}} = 75.5 K_{ep}\left(\frac{1 ~{\rm
cm}^{-3}}{N_g^{SCR}}\right)
\left(\frac{10^5 ~{\rm yr}}{T_{SN}}\right)
\left(\frac{10~ \mu {\rm G}}{B}\right)^2.
\end{equation}
It is independent of the \gr energy for $\epsilon_{\gamma} >
\epsilon_{\gamma}^{*}$, and only given by eq. (25) with
$N_{SN}^e/N_{SN}=1$ for $\epsilon_{\gamma} < \epsilon_{\gamma}^{*}$. This
consideration shows that for the parameters assumed, and for $K_{ep}$ of
the order of $10^{-2}$, we have an IC contribution to the average \gr
background which is comparable to the hadronic background for all energies
above a few GeV.

\subsubsection{Bremsstrahlung contribution}
At high energies, $\epsilon_e, \epsilon_{\gamma} \gg m_e {\rm c}^2$, we
have for the Bremsstrahlung \gr emissivity $Q_{\gamma}^{Br}$
\begin{equation}
Q_{\gamma}^{Br}(\epsilon_{\gamma})=2\int_{\epsilon_{min}}^{\infty} d\epsilon_e
\frac{d\sigma_{ep}^{Br}}{d\epsilon_{\gamma}}cN_g ^{SCR}n^e_{SCR}(\epsilon_e),
\end{equation}
where $\epsilon_{min}$ is the minimum electron energy necessary to produce
a Bremsstrahlung \gr of energy $\epsilon_{\gamma}$, the factor 2 takes
into account the contributions of electron-electron and electron-proton
collisions, and where the differential electron-proton Bremsstrahlung
cross-section is given by (e.g. Berezinsky et al. 1990)
\begin{equation}
\frac{d{\sigma}^{Br}(\epsilon_e,\epsilon_{\gamma})}{d\epsilon_{\gamma}}=
\frac{4{\alpha}r_0^2}{\epsilon_{\gamma}}
\left[\frac{4}{3}-\frac{4}{3}\frac{\epsilon_{\gamma}}{\epsilon_e}+
\left(\frac{\epsilon_{\gamma}}{\epsilon_e}\right)^2\right]
\left[{\rm
ln}\left(\frac{2\epsilon_e}{mc^2}\frac{\epsilon_e-\epsilon_{\gamma}}
{\epsilon_{\gamma}}\right)-\frac{1}{2}\right].
\end{equation}
Here $\alpha\approx 1/137$ and $r_0= 2.818\times 10^{-13}$~cm denote the
fine structure constant and the classical electron radius, respectively.

For our chosen value ${\gamma}_{SCR}=2$, the integral for
$Q_{\gamma}^{Br}(\epsilon_{\gamma})$ can be calculated in closed form. In
the limit ${\rm ln}(\epsilon_{\gamma}/ m_e c^2) \gg 1$, of interest here,
it reduces to the asymptotic form
\begin{equation}
Q_{\gamma}^{Br} (\epsilon_{\gamma})=8 \alpha r_0^2 
c N_g ^{SCR}n^e_{SCR} ( \epsilon_{ \gamma})
{\rm ln}\left( \frac{ \epsilon_{\gamma}}{ m_e c^2}\right).
\end{equation}
Thus, finally, we obtain
\begin{equation}
\frac{Q_{\gamma}^{Br}}{Q_{\gamma}^{SCR}}
= \frac{8 {\alpha} r_0^2 K_{ep}}{Z_{\gamma}^{SCR}{\sigma_{pp}}}
\frac{N_{SN}^e}{N_{SN}}
{\rm ln} \left(\frac{\epsilon_{\gamma}}{m_ec^2}\right)=
9.3 K_{ep} \frac{N_{SN}^e}{N_{SN}}
\left[1+0.066 {\rm ln}\left(\frac{\epsilon_{\gamma}}{1~\mbox{TeV}}\right)\right]
\end{equation}
This small ratio implies that Bremsstrahlung \grs play no role for the
average \gr background above GeV energies, if $K_{ep}\ll 0.1$, taking into
account, that $N_{SN}^e/N_{SN}$ is always smaller than 1.

\section{Discussion}
We note that of order ten SNRs of age younger than $10^5$~yr can on average
lie within a 1 degree field of view of a detector directed towards the
Galactic Center. Therefore a moderate fluctuation of the measured \gr
intensity is expected due to variations of the actual number of SNRs
within such a detector's field of view. At the same time, for directions
perpendicular to the Galactic plane, the chance to observe the
contribution of SCR \gr emission is quite negligible. A question is then
how one might best study the nonuniformities of this background
experimentally. Clearly this is an investigation of its own. Therefore we
would like to restrict ourselves to a few comments here. Due to the
spectral form of the background its graininess is most pronounced at high
\gr energies. This is even more true due to the fact that for individual
sources with a very low magnetic field the IC emission could be much
stronger than the \gr emission due to hadronic interactions; in addition
the IC emission has a harder spectrum. This suggests the use of imaging
atmospheric Cherenkov telescopes. Their resolution in angle and energy is
as good or better than that of other ground-based detectors. However, the
study of extended sources is not an easy task with imaging telescopes
which have a very limited field of view, even employing the stereoscopic
method. For low brightness extended sources a satellite instrument like
the future GLAST detector is well suited since it does not have to deal
with the charged CR background due to its use of an anticoincidence
shield. On the other hand, the statistics achievable with a small area
space detector gets very low above a few tens of GeV. Thus one should
probably attempt such a study with both types of instruments due to their
complementary properties.

A different question concerns the limitations of our approach due to its
concentration on SNRs as the sources of the GCRs. In fact the
considerations in this paper can be applied to any alternative class of
dominant GCR sources. The most important aspect, which leads to the
dominance of SCRs in high-energy \gr production, is that the GCR sources
should generate SCRs with a spectrum that is significantly harder than the
GCR spectrum. Eq. (15) is valid for an arbitrary class of CR sources if we
substitute some other value of the SCR confinement time $T_S$ instead of
$T_{SN}$, since the grammage $x$ is an experimentally fixed quantity. Let
us then assume that some class of compact CR sources produces an energy
$E_C$ in the form of CRs with spectrum $N_{SCR}\propto
\epsilon^{-\gamma_{SCR}}$ with average frequency $\nu_S$, and let us
further assume that this spectrum remains unchanged inside the source
regions for some period of time $T_S$ after which these CRs are released
into the ISM as the GCRs. It is obvious that due to the general energy
requirement the production rate $\nu_S E_C$ should be about the same as
$\nu_{SN}\delta E_{SN}$. The SCR energy $E_C$, deposited in some initial
volume $V$, will produce a dynamically significant disturbance in the
background ISM if we assume that the initial SCR energy density $E_C/V$ is
much greater than the thermal ISM energy density which in turn is of order
$e_{GCR}$. This will inevitably lead to the confinement of these SCRs
inside an expanding, disturbed volume for some period of time $T_S$ before
the SCRs will be released. It is difficult to give a general relation
between $E_C$ and $T_S$ and there may exist only lower bounds on $T_S$,
given $E_C$. Therefore, we cannot exclude speculative source classes with
many weak but long-lived sources. The opposite case of many weak and
short-lived sources is excluded to the extent that the present explanation
of the hard \gr spectrum by the contribution of the SCRs is unique.

\section{Summary}
Our considerations suggest that the SCRs can provide an essential
contribution to the high-energy Galactic \gr flux. According to our
estimates, depending on the parameters, the SCR contribution is less than
10\% of the GCR contribution at GeV energies and it dominates at energies
greater than 100~GeV due to its essentially harder spectrum. This
conclusion is confirmed by calculations performed for the case when SNRs
are the main source of GCRs. At TeV energies the SCRs increase the
expected \gr flux from the Galactic disk by almost an order of magnitude.

The single physical parameter which determines the SCR contribution due to
hadronic interactions is the SCR confinement time $T_{SN}$. As far as the
\gr emission due to $\pi_0$-decay is concerned, the above conclusions are
valid for $T_{SN}\sim 10^5$~yr. Since this SCR contribution is
proportional to $T_{SN}$, it would be negligible at TeV energies if
$T_{SN}\lsim 10^4$~yr. A SNR age of $10^4$~yr typically corresponds to the
intermediate Sedov phase, when the SNR shock is still quite strong.
Therefore it seems to be quite improbable that the GCRs are replenished
from SNRs at such an early phase. For the IC contribution even a ten times
shorter source life time would be sufficient at TeV energies. In fact, for
the TeV IC emission the relevant time scale is the life time
$\tau_e(\epsilon_e)$ of parent SCR electrons due to their synchrotron
losses, which is indeed about $ 10^4$~yr. For decreasing \gr energies
$\tau_e(\epsilon_e)$ increases beyond $10^4$~yrs, and therefore a source
life time of this magnitude would become a limiting factor.

Our conclusions remain valid for alternative classes of possible GCR
sources with comparable overall energy release and comparable individual
confinement times. We note that this contribution of the dominant GCR
sources necessarily exists. As we argue, it may be sufficient by itself to
explain the observed \gr excess, at least in the inner Galaxy where it is
well documented, without a need to invoke additional particle populations
(e.g. Pohl \& Schlickeiser 1991).

\noindent {\bf Acknowledgements}~~
This work has been supported in part by the Russian Foundation of Basic
Research, grants 97-02-16132 and 00-02-17728, and through the Verbundforschung
Astronomie/Astrophysik of the German BMBF, grant 05-2HD66A(7). EGB
acknowledges the hospitality of the Max-Plank-Institut f\"ur Kernphysik
where part of this work was carried out. HJV in turn acknowledges the
hospitality of the Institute of Cosmophysical Research and Aeronomy 
where the final
parts of this work were done. The authors thank the anonymous referee for
directing their attention to the possible role of electron Bremsstrahlung
emission.


\end{document}